\documentclass[]{spie}  
\usepackage{amsbsy}

\title{A theory of magnetization reversal in nanowires}

\author{Robert S. Maier
\skiplinehalf
Depts.\ of Mathematics and Physics, University of Arizona,
Tucson, AZ 85721, USA \\
}

\authorinfo{Further author information: \\E-mail:
rsm@math.arizona.edu, Telephone: +1 520 621 2671}

\begin{document} 
\maketitle 

\begin{abstract}
Magnetization reversal in a ferromagnetic nanowire which is much narrower
than the exchange length is believed to be accomplished through the
thermally activated growth of a spatially localized nucleus, which
initially occupies a small fraction of the total volume.  To~date, the most
detailed theoretical treatments of reversal as a field-induced but
noise-activated process have focused on the case of a very long
ferromagnetic nanowire, i.e., a highly elongated cylindrical particle, and
have yielded a reversal rate per unit length, due to an underlying
assumption that the nucleus may form anywhere along the wire.  But in a
bounded-length (though long) cylindrical particle with flat ends, it is
energetically favored for nucleation to begin at either end.  We~indicate
how to compute analytically the energy of the critical nucleus associated
with either end, i.e., the activation barrier to magnetization reversal,
which governs the reversal rate in the low-temperature (Kramers) limit.
Our treatment employs elliptic functions, and is partly analytic rather
than numerical.  We also comment on the Kramers prefactor, which for this
reversal pathway does not scale linearly as the particle length increases,
and tends to a constant in the low-temperature limit.
\end{abstract}



\section{INTRODUCTION}
\label{sec:intro}

The phenomenon of magnetization reversal is of intense interest to the data
storage and recording industries, for obvious reasons.  Most treatments of
this phenomenon have employed the standard techniques of micromagnetic
modeling: analytically or numerically solving a partial differential
equation satisfied by the magnetization vector, which is valid in an
averaged, coarse-grained sense, on scales large compared to the crystal
lattice~\cite{Fidler2000}.  Ferromagnetic samples of many geometries have
been studied theoretically~\cite{Aharoni2000}.

Provided that a sufficiently strong reversed field is applied to a
ferromagnetic sample, the reversal of the direction of magnetization is an
essentially deterministic process, which if the sample is sufficiently
large (larger than the coherence length) will begin with the formation of a
spatially localized `nucleus' of reversed magnetization.  This nucleus will
grow, eventually yielding a global reversal of the magnetization.  This is
in contrast to the case when the applied field is less strong than the
nucleation field, in which a critical nucleus can only be formed by thermal
fluctuations, though its subsequent growth may be deterministic.  The two
types of reversal are sometimes called field-activated and noise-activated.
In~many applications, such as quantifying the likelihood of loss of stored
data, it is the second phenomenon, more closely tied to statistical
mechanics than to the theory of deterministic evolution equations, which is
of interest.

The difference between the two types of reversal is made clear by an
analogy.  Consider an oversimplified overdamped scalar model of the form
$\dot m=-U'(m)+{\rm noise}$, where $U(\cdot)$~is a bistable effective
potential, the two minima of which correspond to an alignment of the
magnetization with and against an applied field.  As the field strengthens,
this double-well potential may become sufficiently skewed that the latter
of the two minima disappears, leading to a rapid fall into the remaining
minimum, which corresponds to alignment with the field.  This is the
essence of field-activated nucleation.  On the other hand, if the field is
insufficiently strong, then there will remain two minima: a~metastable one
(against the field) and a stable one (with the field).  It~is clear that
noise-induced nucleation is in essence a standard problem of noise-induced
escape from a metastable state.  Of~course, a satisfactory model will need
to be more sophisticated than this scalar model.  Even if the sample is
small (``zero-dimensional''), so that reversal takes place by a process of
spatially coherent rotation of the magnetization vector, the vectorial
nature of the magnetization is crucial.

The process of noise-activated reversal by coherent rotation was first
treated theoretically by Brown~\cite{Brown63}, who by adapting the Kramers
theory of weak-noise escape rates was able to work~out an asymptotic
(low-temperature) expression for the reversal rate in a single-domain
particle with uniaxial anisotropy.  His work has been extended to the case
of cubic anisotropy.  However, the issue of which reversal mechanisms are
dominant in {\em large\/} ferromagnetic particles remains controversial.
The simplest sort of non-zero-dimensional sample is a ferromagnetic
nanowire: a~long, thin cylinder of width at~most a few tens of nanometers.
The most extensive treatment of field-induced but noise-activated
magnetization reversal in such a sample is that of
Braun~\cite{Braun93,Braun94a,Braun94b,Braun94c,Braun94d}.  By applying the
infinite-dimensional Kramers theory of Langer and
Coleman~\cite{Schulman81chapter29} to a stochastically perturbed
Landau--Lifshitz--Gilbert equation governing the evolution of the
magnetization vector along the nanowire, he worked~out analytically an
asymptotic expression for the weak-noise reversal rate.  However, his
treatment assumes a {\em toroidal\/} sample (with periodic boundary
conditions), and applies only in the limit of infinite sample length.  In
this limit, the nucleus of his model is a metastable field configuration of
soliton type, with a simple functional form.  It comprises a `critical
droplet' of reversed magnetization bounded by a $\pi$~wall and a~$-\pi$
wall; this droplet may form with equal likelihood anywhere along the
sample.  His computed reversal rate is accordingly proportional to the
sample length~$L$.  It~also contains a factor $\exp(-\Delta E/k_BT)$, where
$\Delta E$~is the energy of formation of the droplet.  Like Brown's
asymptotic reversal rate formula, Braun's contains corrections to the
Arrhenius (i.e., exponential) low-temperature falloff.  For reasons
explained below, both fall~off as~$T^{-1/2}\exp(-\Delta E/k_B T)$
as~$T\to0$.  A~non-Arrhenius behavior of the reversal rate as~$T\to0$ has
been found in stochastic simulations of single-domain particles,
at~least~\cite{Boerner98}.

Aharoni~\cite{Aharoni95,Aharoni96} has criticized Braun's not including the
(nonlocal) magnetostatic self-energy in his Landau--Lifshitz--Gilbert
Hamiltonian.  If the magnetization is not solenoidal
(${\boldsymbol\nabla}\cdot{\bf m}\neq0$), there will be a magnetostatic
volume charge, which will generate a demagnetizing field.  This term can be
neglected in a planar topology~\cite{Broz90}, but not in the cylindrical
topology.  Aharoni has also criticized the assumption that the
magnetization vector is constant across the cross-section of the
cylindrical sample, which prevents reversal from occurring via a `curling'
mode.  However, this assumption is presumably justified in the limit of
small sample width.  Also, by what amounts to dimensional analysis, Braun
has shown that in this limit the nonlocal part of the self-energy is
relatively small, and the local part may be absorbed into a renormalization
of crystalline anisotropy constants~\cite{Braun99}.  With these
emendations, Braun's analysis seems to be applicable to thin cylindrical
samples of at~least some materials, though of~course the assumption of a
very long sample with periodic boundary conditions is unrealistic and will
need to be dropped.

In this extended abstract we indicate how the analytic treatment of Braun
can be extended to the case of a thin cylinder with flat ends, at which
open (i.e., Neumann) boundary conditions on the magnetization vector are
appropriate.  We~do this by adapting our work on reversal in a real
Ginzburg--Landau equation perturbed by weak spatiotemporal
noise~\cite{Maier02}.  Since we do not take the $L\to\infty$ limit, our
expression for the metastable nucleus configuration is more complicated
than Braun's.  It~involves elliptic functions~\cite{Abramowitz65}, which
have long been used in the analytic modeling of periodic micromagnetic
structures.  (See ref.~\citenum{Brown63a},~\S2.4, and
ref.~\citenum{Aharoni60}.)  Sophisticated numerical simulations at the
magnetic moment level of noise-activated reversal, in a thin cylinder with
flat ends, show that a droplet of reversed magnetization forms at one of
the ends and spreads inward~\cite{Brown00,Brown2001}.  This reversal mode
is in agreement with our analytic treatment.  We~also comment on the
prefactor in the low-temperature reversal rate, i.e., the factor that
multiplies $\exp(-\Delta E/k_B T)$.  As~$L\to\infty$, it does not grow
linearly in~$L$, but rather tends to a constant, since the formation of a
droplet of reversed magnetization away from both ends of the cylinder is
energetically disfavored.  We~also find that the prefactor includes
no~$T^{-1/2}$ factor.  That~is, the rate of noise-activated magnetization
reversal, if physically reasonable boundary conditions are imposed, follows
an Arrhenius law.

\section{ESCAPE FROM A METASTABLE STATE}
\label{sec:kramers}

The theory of noise-activated fluctuations away from a metastable
state~\cite{Hanggi90}, including escape to a stable state, was developed
most significantly by Kramers~\cite{Kramers40}.  Major contributions were
also made by Eyring, who worked on escape in multidimensional state spaces
in the context of reaction-rate theory.  It~should be emphasized that the
Kramers approach to escape rates is {\em asymptotic\/}: increasingly
accurate in the weak-noise limit, in which escape per unit time becomes
exponentially unlikely.  In this limit, the phenomenology of
noise-activated escape from the domain of attraction of a metastable state
involves eventual motion along an `optimal trajectory' that typically goes
uphill from the metastable state to a saddle point, near which exit occurs.
By definition, this is the escape pathway of least resistance.  The escape
rate is computed as the integral of the outgoing probability flux through a
region on the boundary of the domain of attraction which is centered on the
saddle point.  The optimal trajectory picture and the associated
computations become increasingly valid in the weak-noise limit.

The present author and his collaborators have extended Kramers theory to
the case when the system being modeled, if noise is absent, has
`nongradient' dynamics: for example, when its order parameter~${\bf x}$
(not a magnetization, in~general) is finite-dimensional but does not evolve
according to any law of the form $\dot{\bf x}=-{\boldsymbol\nabla}U({\bf
x})$.  This could be because the dynamics include dissipation or
non-conservative forces.  In the nongradient case the phenomenology of
escape may be quite different from that described above.  The domain of
attraction of a metastable state may have an unstable limit cycle as its
boundary, and the outgoing optimal trajectory may spiral into this limit
cycle.  This is in~fact a feature of the abovementioned coherent rotation
model of Brown~\cite{Brown63}, which is Hamiltonian but includes a damping
term.  In two-dimensional models in which the boundary of the domain of
attraction is an unstable limit cycle, the Kramers escape rate must be
computed as the integral of the outgoing probability flux over the entire
boundary, rather than merely over the vicinity of a saddle point (which may
not be present).  This usually gives rise to anomalous weak-noise behavior
of the Kramers prefactor, i.e., the factor that multiplies the usual
exponential suppression factor in the asymptotic weak-noise escape rate.

In this section we review the extensions to Kramers rate theory that we
shall need in understanding and extending the results of
Braun~\cite{Braun93,Braun94a,Braun94b,Braun94c,Braun94d}, with an emphasis
on the prefactor and on the phenomenology of escape.  We~shall use
dimensionless units throughout.  Suppose a point~$\bf x$ in
finite-dimensional Euclidean space (or~possibly on a surface such as a
sphere) evolves according to the overdamped Langevin equation
\begin{equation}
\label{eq:Langevin}
\dot{\bf x}={\bf u}({\bf x}) + \epsilon^{1/2}{\boldsymbol\eta}(t).
\end{equation}
Here $\bf u$ is a drift field, ${\boldsymbol\eta}$~is normalized white noise,
which satisfies
\begin{equation}
\langle\eta_i(t_1)\eta_j(t_2)\rangle = \delta_{ij}\delta(t_1-t_2),
\end{equation}
and $\epsilon\ll1$~is a noise-strength parameter.  Suppose the drift field
has a metastable point~$S$ and a stable point~$S'$, with a saddle point
denoted~$U$ (for unstable) between them, on the common boundary of their
domains of attraction.  If~${\bf x}=S$ initially, the time for
noise-activated escape can be defined as the time needed for $\bf x$ to
reach some specified region~$R$ that lies strictly within the domain of
attraction of~$S'$, and contains $S'$ itself.  This escape time $t_{\rm
exit}(S\to S')$ is a random variable, the distribution of which in the
weak-noise limit will not be much affected by the choice of region~$R$.  As
$\epsilon\to0$, its mean $\langle t_{\rm exit}(S\to S')\rangle$ will grow
exponentially (to~leading order), and its distribution will increasingly
resemble that of an exponential random variable.  The escape rate
$\Gamma_{S\to S'}$ can be defined as the reciprocal $\langle t_{\rm
exit}(S\to S')\rangle^{-1}$.  Just as there is noise-activated motion from
$S$ to~$S'$, there is noise-activated motion from $S'$ back to~$S$, but
reverse motion is exponentially suppressed in the weak-noise limit,
relatively as~well as absolutely.

The escape rate $\Gamma_{S\to S'}$ may be written as $\Gamma^0\exp(-\Delta
W/\epsilon)$, where $\Delta W$~is the minimum amount of energy the noise
must put into the system to drive it over the saddle~$U$.  By definition,
the second factor is the Kramers factor, and $\Gamma^0$~is the Kramers
prefactor.  The case when the drift field~$\bf u$ is the negative gradient
of a potential function, i.e., ${\bf u}=-{\boldsymbol\nabla} V$, is the
easiest case to treat, since the optimal escape trajectory is simply a
time-reversed relaxational trajectory, and satisfies $\dot{\bf x}=-{\bf
u}({\bf x})$.  It turns~out that $W=2V$ (the factor of~$2$ is sometimes
absorbed into a redefinition of~$\epsilon$).  The weak-noise behavior of
the escape rate can readily be expressed in~terms of linear approximations
to the drift field at ${\bf x}=S$ and~${\bf x}=U$, by the Kramers--Eyring
formula.  In the two-dimensional case, this formula is
\begin{equation}
\label{eq:Eyring}
\Gamma_{S\to S'}\sim
\frac1{2\pi}\sqrt{\left|\lambda_{\parallel}(U)\right|\lambda_{\parallel}(S)}
\sqrt{\lambda_{\perp}(S)\over \lambda_{\perp}(U)}\exp(-2\Delta V/\epsilon)\,,\qquad\epsilon\to0\,.
\end{equation}
Here $\lambda_\parallel(S)\equiv-{\partial u_{S,\parallel}/\partial
x_{S,\parallel}}$ is the eigenvalue of the linearized negative drift field
at~$S$ whose corresponding eigenvector points along $\hat {\bf
x}_{S,\parallel}$, the direction locally parallel to the departing optimal
trajectory\null.  Similarly, $\lambda_\perp(U)\equiv-{\partial
u_{U,\perp}/\partial x_{U,\perp}}$ is the eigenvalue of the linearized
negative drift field at~$U$ whose corresponding eigenvector points
along~$\hat {\bf x}_{U,\perp}$, the direction locally perpendicular to the
approaching optimal trajectory, and so forth.  $\lambda_\parallel(U)$~is
the eigenvalue of the linearized negative drift field at~$U$ that
corresponds to the unstable, or `downhill' direction; it is negative.  The
formula~(\ref{eq:Eyring}) generalizes immediately to the case when the
order parameter $\bf x$~has arbitrary finite dimensionality~$d$.
In~general, the prefactor will contain a product of $d-1$ factors
resembling $\sqrt{\lambda_{\perp}(S)\over \lambda_{\perp}(U)}$, one for
each direction transverse to the optimal trajectory.

The standard formula~(\ref{eq:Eyring}) contains a constant prefactor, which
does not include any power of the noise strength~$\epsilon$.  But the
derivation of~(\ref{eq:Eyring}) makes clear how such a power could arise.
The {\em stationary\/} probability density function~(PDF) of the order
parameter~$\bf x$ is $\rho_0({\bf x})\equiv\exp(-W({\bf x})/\epsilon)$,
up~to normalization.  The slowest decaying nonstationary mode of the
drift-diffusion equation for the PDF is a {\em quasistationary\/}
PDF~$\rho_1$, which characterizes how probability equilibrates between the
domains of attraction of $S$ and~$S'$.  Like~$\rho_0$, it will contain an
$\exp\left(-2V({\bf x})/\epsilon\right)$ factor, so it will typically be
concentrated on the $O(\epsilon^{1/2})$ transverse lengthscale near the
saddle point~$U$.  But unlike~$\rho_0$, $\rho_1$~will have opposite signs
in the two domains.  Along the $\hat{\bf x}_{U,\parallel}$ direction
through~$U$, it will change sign.  The rate $\Gamma_{S\to S'}$ is the rate
at which probability diffuses across the saddle, from $S$ toward~$S'$.  As
Kramers showed, this is easily evaluated as an integral of the probability
flux.  In this integral there are several powers of~$\epsilon$ that must be
kept track~of.  For example, there is a normalization factor proportional
to~${\epsilon}^{-d/2}$ in $\rho_1$ (as~there is in~$\rho_0$).  Also, it
follows from the Langevin equation~(\ref{eq:Langevin}) that besides
drifting along~$\bf u$, the PDF diffuses with diffusivity~$\epsilon$.
A~careful evaluation of the probability flux integral yields the Kramers
prefactor shown in~(\ref{eq:Eyring}), with all powers of~$\epsilon$
canceling.

A prefactor containing a nontrivial power of~$\epsilon$ could arise from
the potential energy~$V=V({\bf x})$ behaving non-quadratically near ${\bf
x}=S$ or~$U$, causing one or more of the four eigenvalues appearing
in~(\ref{eq:Eyring}) to equal zero~\cite{KipNew}.  As mentioned,
non-Arrhenius behavior will also arise in the case of a two-dimensional
order parameter if the separatrix between the domains of attraction of $S$
and~$S'$ contains no~saddle point, and is an unstable limit cycle.  For
this to be the case, the drift~$\bf u$ cannot be the negative gradient of a
potential function.  A~transverse integration of the probability flux
generated by the quasistationary PDF~$\rho_1$ over an $O(1)$ lengthscale
around a limit cycle, rather than over an $O(\epsilon^{1/2})$ transverse
lengthscale near a saddle point, will yield an escape rate larger by a
factor roughly proportional to~$\epsilon^{1/2}$.  In~such a situation,
\begin{equation}
\label{eq:needed}
\Gamma_{S\to S'}\sim {\rm const}\times\epsilon^{-1/2}\exp(-\Delta
W/\epsilon)\,,\qquad\epsilon\to0\,,
\end{equation}
where `${\rm const}$' can be evaluated as an integral around the
separatrix, will replace~(\ref{eq:Eyring}).

All of the preceding remarks apply to the case of finite-dimensional {\em
overdamped\/} dynamics, where the Langevin equation for the order parameter
has the form~(\ref{eq:Langevin}).  If~there is damping but the strength of
the damping is not taken to infinity, a more appropriate Langevin equation
in dimensionless units is
\begin{equation}
\label{eq:nonoverdamped}
\ddot {\bf x} + \alpha\dot {\bf x} = {\bf u}({\bf x}) + \epsilon^{1/2} {\boldsymbol\eta}(t),
\end{equation}
where $\alpha>0$~is a damping constant, or equivalently
\begin{equation}
\left\{
\begin{array}{ccl}
\dot{\bf x} &=& {\bf v},\\
\dot{\bf v} &=& -\alpha{\bf v}+{\bf u}({\bf x}) + \epsilon^{1/2}{\boldsymbol\eta}(t),
\end{array}
\right.
\end{equation}
where ${\bf v}\equiv\dot{\bf x}$.  This is effectively an overdamped
Langevin equation for a $2d$-dimensional order parameter~$({\bf x},{\bf
v})$ that diffuses only in the velocity direction.  But if $\alpha$~has
some fixed positive value, the weak-noise behavior of the Kramers prefactor
in this non-overdamped model is surprisingly similar to that discussed
above.  The prefactor becomes $\alpha$-dependent~\cite{Hanggi90}, but there
is still no $\epsilon$-dependence if ${\bf u}=-{\boldsymbol\nabla} V$ with
$V$~quadratic near $S$ and~$U$.  When $d=2$, an $\epsilon^{-1/2}$ factor
will appear in the Kramers prefactor if ${\bf u}$~is not derived from a
potential and there is no~saddle on the separatrix, for essentially the
reason already discussed.  There are clearly some subtleties associated
with the reduction of the non-overdamped model of~(\ref{eq:nonoverdamped})
to the overdamped model of~(\ref{eq:Langevin}) when~$\alpha\to\infty$, but
we do not discuss them here.

Infinite-dimensional Kramers theory is a theory of rate and relaxation
phenomena exhibited by random fields, rather than by randomly evolving
finite-dimensional order parameters.  However, it is formally quite similar
to finite-dimensional Kramers theory.  This is perhaps not entirely clear
from standard expositions~\cite{Schulman81chapter29}, but we shall make
it~so.  For simplicity, consider a field $\phi=\phi(x,t)$ perturbed by
noise of strength~$\epsilon$, which evolves with overdamped dynamics on a
spatial domain~$[0,L]$ according to
\begin{equation}
\dot\phi = -\delta{\cal H}/\delta\phi + \epsilon^{1/2}\eta(x,t).
\end{equation}
The first term on the right-hand side is the functional derivative of an
energy functional, and $\eta(x,t)$~is normalized {\em spatiotemporal\/}
white noise, which satisfies
\begin{equation}
\label{eq:standard}
\langle \eta(x_1,t_1) \eta(x_2,t_2) \rangle 
= \delta(x_1-x_2)\delta(t_1-t_2),
\end{equation}
i.e., has a flat power spectrum in space as~well as time.  Typically
\begin{equation}
{\cal H}[\phi] = \int_0^L \left[
\frac{\phi'(x)^2}2 + V\left(\phi(x)\right)
\right]\,dx,
\end{equation}
so that the evolution equation is a stochastically perturbed
Ginzburg--Landau equation, namely
\begin{equation}
\label{eq:GL}
\dot\phi = \Delta\phi - V'(\phi) + \epsilon^{1/2}\eta(x,t).
\end{equation}
If the potential~$V$ is bistable or multistable then noise-induced
transitions between stable spatially extended states, each of which is
concentrated around a field value that is a minimum of~$V$, may
occasionally take place.  Each such transition will involve a passage
through an intermediate (unstable) saddle state.

These spatially extended states are strongly affected by the boundary
conditions at $x=0$ and~$x=L$.  Suppose that $V(\phi)$ has two local
minima, at $\phi=S,S'$, with the latter being the global minimizer; and
that the intervening local maximum is at~$\phi=U$.  If~periodic or Neumann
boundary conditions are imposed, one would expect that the metastable,
saddle, and stable states are the constant field configurations $\phi\equiv
S,U,S'$; though if (say) Dirichlet boundary conditions are imposed, one
would expect them to differ, especially near $x=0$ and~$x=L$.  Actually,
even in the periodic and Neumann cases the saddle state most relevant to
noise-induced transitions may not be the constant solution~$\phi\equiv U$.
If $L$~is sufficiently large, there may be a lower-energy saddle in the
field configuration space, which is spatially oscillatory rather than
constant~\cite{Maier02}.  The importance of such nonconstant saddle
configurations was first pointed~out by Langer~\cite{Langer67} in a study
of the $L\to\infty$ limit of escape from a metastable state.  Due to their
oscillatory character they are sometimes called ``Langer's bounces''.  Each
such oscillatory saddle configuration can be viewed as a background
metastable ($\phi\equiv S$) state, within which a droplet of the stable
($\phi\equiv S'$) state has formed.  The droplet is critical, which means
that if pushed in the right direction, the saddle configuration may roll
`downhill' to the stable rather than the original metastable state.

The Kramers--Eyring formula~(\ref{eq:Eyring}) generalizes readily to the
infinite-dimensional case, becoming
\begin{equation}
\label{eq:infdem}
\Gamma\sim\frac{\left|\lambda_0(\phi_u)\right|}{2\pi}
\sqrt{
\frac
{\prod_{n=0}^\infty
\lambda_n(\phi_s)}
{\prod_{n=0}^\infty\left|\lambda_n(\phi_u)\right|}
}
\,\exp(-2\Delta{\cal H}/\epsilon),\qquad \epsilon\to0
\,.
\end{equation}
Here $\{\lambda_n(\phi_s)\}_{n=0}^\infty$ and
$\{\lambda_n(\phi_u)\}_{n=0}^\infty$ are the eigenvalues of the
linearizations of the negative drift field $\delta{\cal H}/\delta\phi$ at
the metastable state $\phi=\phi_s(x)$ and the dominant saddle
$\phi=\phi_u(x)$, arranged in increasing order.  The only negative
eigenvalue in either set is~$\lambda_0(\phi_u)$, which corresponds to
`downhill' motion away from the saddle.  It~is easy to see that in the
two-dimensional case, (\ref{eq:infdem})~reduces to~(\ref{eq:Eyring}).

Because the numerator and denominator of the quotient inside the square
root are products of an infinite number of eigenvalues, which tend to
infinity as~$n\to\infty$, each diverges.  However, their quotient can be
defined as a limit of finite-dimensional truncations.  Usually it is recast
as a {\em determinant quotient\/}, computed from the linearized
deterministic evolution operators at
$\phi=\phi_s,\phi_u$~\cite{Langer69,McKane95}.  Consider for example a
small perturbation~$\xi$ away from the metastable state, i.e.,
$\phi=\phi_s+\xi$.  Then to leading order
$\dot\xi=-\hat\Lambda[\phi_s]\xi$, where
$\hat\Lambda[\phi_s]\equiv\delta^2{\cal H}/\delta\phi^2[\phi_s]$ specifies
the linearized zero-noise dynamics near~$\phi=\phi_s$.  Similarly,
$\hat\Lambda[\phi_u]$ specifies the dynamics near~$\phi=\phi_u$.  For the
model defined by~(\ref{eq:standard}),
$\hat\Lambda[\phi_s],\allowbreak\hat\Lambda[\phi_u]$ equal
$-\Delta+V''\left(\phi_s(x)\right),\allowbreak-\Delta+V''\left(\phi_u(x)\right)$.
One can write
\begin{equation}
\label{eq:asymp1}
\Gamma_{\phi_s\to\phi_{s'}}\sim\frac{\left|\lambda_0(\phi_u)\right|}{2\pi}
\sqrt{
\frac{\det{\hat\Lambda[\phi_s]}}
{|\det{\hat\Lambda[\phi_u]}|}
}
\,\exp(-2\Delta{\cal H}/\epsilon),\qquad \epsilon\to0
\,.
\end{equation}
The two determinants of Schr\"odinger operators are not defined
individually, but their quotient can be made sense of by any of several
regularization schemes; for example, zeta-function and dimensional
regularization~\cite{McKane95}.

If the boundary conditions at $x=0,L$ are periodic, so that the spatial
domain~$[0,L]$ is effectively a circle, and the dominant (lowest-energy)
saddle state $\phi=\phi_u(x)$ is not the constant solution $\phi\equiv U$
but rather a field configuration containing a critical droplet, the
asymptotic formula~(\ref{eq:asymp1}) must be modified.  Since the droplet
configuration can be shifted arbitrarily around~$[0,L]$, the operator
$\hat\Lambda[\phi_u]$ will have a zero eigenvalue~$\lambda_1(\phi_u)$, the
accompanying eigenfunction being a Goldstone mode that expresses
translation invariance.  So in this case the prefactor of~(\ref{eq:asymp1})
will diverge, signalling a change in the phenomenology of escape.  In~the
infinite-dimensional space of field configurations, the separatrix between
the domains of attraction of the metastable and stable states will now
include an {\em unstable limit cycle\/} of length~$L$, and noise-activated
escape may take place anywhere along~it.  We~have already explained how to
handle such separatrices.  Since the escape rate must be computed by
integrating the outgoing probability flux around the separatrix, rather
than on a transverse lengthscale of magnitude~$O(\epsilon^{1/2})$, the
prefactor will be larger by a factor that grows like~$L\epsilon^{-1/2}$
as~$L\to\infty$.  The asymptotic weak-noise behavior of the escape rate in
this case, when $L$~is fixed, is accordingly of the form
\begin{equation}
\label{eq:novel}
\begin{array}{ll}
\Gamma_{\phi_s\to\phi_{s'}}\,\sim\, C(L)\,\epsilon^{-1/2} \exp\bigl(-2(\Delta{\cal H})(L)/\epsilon\bigr),&\quad\epsilon\to0\,,\\[\jot]
C(L)\,\sim\,{\rm const}\times L,&\quad L\to\infty\,,
\end{array}
\end{equation}
where $2(\Delta{\cal H})(L)$ is the energy barrier, i.e., the energy of
formation of the critical droplet within the spatial domain~$[0,L]$.  This
will tend to a limiting value as~$L\to\infty$, since the droplet tends to a
limiting shape.  We~stress that the asymptotic linear growth of the
prefactor with~$L$ in this case occurs for a simple physical reason:
if~periodic boundary conditions are imposed, the critical droplet may form
with equal likelihood anywhere in the spatial domain.  Departures from
linearity will occur when $L$~is small, but they are due~to the droplet
assuming its limiting shape only in the large-$L$ limit.  The large-$L$
behavior of the prefactor displayed in~(\ref{eq:novel}) is well-known, but
the finite-volume case of fixed~$L$ has only recently been systematically
treated~\cite{Maier02}.

\section{STOCHASTIC MICROMAGNETISM}
\label{sec:micromagnetism}

We can now relate the low-temperature micromagnetic reversal rate theories
of Brown~\cite{Brown63} and
Braun~\cite{Braun93,Braun94a,Braun94b,Braun94c,Braun94d} to
finite-dimensional and infinite-dimensional Kramers theory, respectively,
and explain the physical origin of the $T^{-1/2}$ prefactor that modifies
each of their Arrhenius laws.  Our discussion will indicate how Braun's
treatment of a very long ($L\to\infty$) ferromagnetic nanowire can be
modified and extended to the case of a thin ferromagnetic cylinder with
flat ends, and fixed length~$L$.

The starting point for continuum micromagnetic modeling~\cite{Fidler2000}
is the Landau--Lifshitz--Gilbert equation for the bulk magnetization
vector~$\bf m$.  If dimensionless units are used and the magnetization unit
is chosen so that the saturation magnetization of the ferromagnetic
material being modeled equals unity, i.e., so that $|{\bf m}|=1$ at each
point in the sample, this equation is
\begin{equation}
\label{eq:llg1}
\dot{\bf m}= -\gamma {\bf m}\times{\bf h}_{\rm eff} + \alpha {\bf
m}\times\dot{\bf m},
\end{equation}
where $\gamma>0$ is the gyromagnetic ratio.  The first term causes $\bf m$
to precess about the local magnetic field~${\bf h}_{\rm eff}$.  The second
is a phenomenological damping term introduced by Gilbert~\cite{Gilbert55},
which causes $\bf m$ to relax toward the direction of~${\bf h}_{\rm eff}$.
This becomes clear if the equation is rewritten as
\begin{equation}
\label{eq:llg2}
(1+\alpha^2)\gamma^{-1}\dot{\bf m} = -{\bf m}\times{\bf h}_{\rm eff} -
\alpha{\bf m}\times({\bf m}\times{\bf h}_{\rm eff}),
\end{equation}
which follows by evaluating the cross product of~$\bf m$ with each term.
The Gilbert damping term, which is now universally used as a replacement
for a less satisfactory damping term proposed earlier by Landau and
Lifshitz~\cite{Iida63}, is still slightly controversial, in~part because of
the difficulty of calculating the material-dependent dimensionless damping
coefficient $\alpha>0$ from first principles.

The formula for~${\bf h}_{\rm eff}$ must capture much of the physics.  If
the dynamics are essentially Hamiltonian (with the exception of the Gilbert
damping), then ${\bf h}_{\rm eff}=-\delta H[{\bf m}]/\delta{\bf m}$ for
some Hamiltonian functional $H[{\bf m}]$ of the magnetic field
configuration.  The Hamiltonian employed by Braun for a thin cylindrical
particle whose axis extends from $x=0$ to~$x=L$, with ${\bf m}$~taken to
depend only on~$x$, is
\begin{equation}
\label{eq:BraunH}
{\cal H}[{\bf m}] =
\int_{0}^L \left\{
[(\partial_x m_x)^2 + (\partial_x m_y)^2 + (\partial_x m_z)^2]
+ ({\bf m},{\sf K}{\bf m}) - {\bf h}_{\rm ext}\cdot{\bf m}
\right\}\,dx.
\end{equation}
The first term is the exchange term, of quantum-mechanical origin, which
favors the alignment of nearby magnetic moments, and the second is a
quadratic form on the sphere $S^2\ni\bf m$ that models the anisotropy of
the material, if~any.  For convenience the unit of length here is chosen to
equal the exchange length, so the coefficient of the exchange term is
unity.

This Hamiltonian does not include a term of a type arising from classical
electromagnetism, the omission of which has been criticized by
Aharoni~\cite{Aharoni95,Aharoni96}.  If ${\boldsymbol\nabla}\cdot{\bf
m}\neq0$ in the interior of the sample then there will be a magnetostatic
volume charge, which will generate its own magnetic field, with which the
bulk magnetization will interact.  The additional field is called the
`demagnetizing' field, for an obvious reason.  In~general, this $\bf
m$-generated field will add a term to the energy density which is nonlocal
in space.  However, as mentioned in Section~\ref{sec:intro},
Braun~\cite{Braun99} has given convincing arguments for replacing this term
in the thin-cylinder limit by a {\em local\/} term, which can be absorbed
into a renormalization of the anisotropy tensor~$\sf K$.

Braun considered the effects on the Landau--Lifshitz--Gilbert
equation~(\ref{eq:llg2}) of adding a small random term of the form
$\epsilon^{1/2}{\boldsymbol\eta}(x,t)$ to~${\bf h}_{\rm eff}$, where
${\boldsymbol\eta}(x,t)$~is vectorial spatiotemporal white noise.  Before
discussing his results, it is useful to review Brown's much earlier
theoretical treatment of ferromagnetic reversal in a single-domain
particle.  The equation~(\ref{eq:llg2}) makes sense as the evolution
equation of a single vector~$\bf m$, regarded as the spatially uniform
saturation magnetization across such a particle.  The appropriate
Hamiltonian is not~(\ref{eq:BraunH}) but rather
\begin{equation}
{\cal H}[{\bf m}] = ({\bf m},{\sf K}{\bf m}) - {\bf h}_{\rm ext}\cdot{\bf m}\,,
\end{equation}
so
\begin{equation}
{\bf h}_{\rm eff} = -\delta{\cal H}[{\bf m}]/\delta{\bf m} = {\bf h}_{\rm
ext} - 2{\sf K}{\bf m}\,.
\end{equation}
The anisotropy tensor~$\sf K$ should include both shape and crystalline
anisotropies.  Brown considered the case of uniaxial anisotropy, with the
external field directed along the `easy axis' of the particle.  If this
easy axis points in the $x$-direction, ${\sf K}={\rm diag}(-K_e,0,0)$ for
some $K_e>0$, so let ${\bf h}_{\rm ext}=(h_{\rm ext},0,0)$ with $h_{\rm
ext}>0$.  Then as a function of the angle~$\theta$ between $\bf m$~and the
$x$-axis,
\begin{equation}
{\cal H}[{\bf m}] = -K_e \cos^2\theta - h_{\rm ext}\cos\theta.
\end{equation}
Provided that $h_{\rm ext}<2K_e$, this function on the sphere $S^2\ni{\bf
m}$ has two minima: the metastable state $\theta=\pi$ (antiparallel
to~${\bf h}_{\rm ext}$) and the stable state $\theta=0$ (parallel to~${\bf
h}_{\rm ext}$), which has lower energy.  Their energies are respectively
$-K_e\pm h_{\rm ext}$.

It is easy to check that according to the evolution
equation~(\ref{eq:llg2}), the particle magnetization~$\bf m$ will spiral
into one or the other of these two states.  On the sphere, the separatrix
between their domains of attraction is the parallel of latitude singled out
by the equation $\theta=\cos^{-1}(-h_{\rm ext}/2K_e)$.  This is an {\em
unstable limit cycle\/}.  If~a temporal noise term of the form
$\epsilon^{1/2}{\boldsymbol\eta}(t)$, with $\epsilon\ll1$, is added
to~${\bf h}_{\rm eff}$, then the noise will eventually drive the
magnetization vector away from the metastable state and across this
separatrix.  It~can be shown that the optimal trajectory for
noise-activated escape is, in~fact, a time-reversed infalling spiral.  So
the mechanism of activated magnetization reversal in this model is of the
sort discussed in Section~\ref{sec:kramers}.  However, it should be noted
that when $h_{\rm ext}>2K_e$, the metastable state becomes unstable, and
there is no separatrix to be crossed.  Reversal in the presence of a
sufficiently high external field is field-activated rather than
noise-activated.

By examination, the `height' of the unstable limit cycle in~terms of energy
is $h_{\rm ext}^2/4K_e$, so the energy barrier to reversal is $2\Delta{\cal
H}=2\left[h_{\rm ext}^2/4K_e - (-K_e+h_{\rm ext})\right]$.  The weak-noise
reversal rate will be given by~(\ref{eq:needed}), i.e.,
\begin{equation}
\Gamma\sim {\rm const}\times\epsilon^{-1/2}\exp(-2\Delta{\cal
H}/\epsilon)\,,\qquad\epsilon\to0\,.
\end{equation}
The factor `${\rm const}$' can be calculated by integrating along the limit
cycle the rate at which the probability density of~$\bf m$ is absorbed.
However, this derivation already makes it clear why the N\'eel--Brown
reversal rate, as it is now called, falls~off in the limit of low
temperature in a non-Arrhenius way.  The anomalous power $\epsilon^{-1/2}$,
i.e.,~$T^{-1/2}$, arises from the need to integrate a probability flux
around the entire limit cycle, rather than over an $O(\epsilon^{1/2})$
transverse lengthscale near a saddle point, as is more typical in the
Kramers theory of noise-activated escape.  This derivation can be extended
to the case of a single-domain ferromagnetic particle with mildly
non-uniaxial anisotropy tensor~$\sf K$, in which, as here, the barrier to
reversal is an unstable limit cycle.

We now turn to the thin nanowire computations of Braun.  In their full
generality, they apply to a thin nanowire of a ferromagnetic material with
cubic anisotropy, the easy axis of which is aligned with the axis of the
wire; which may be taken to be the $x$-axis.  By adding an appropriate
multiple of the identity matrix to~$\sf K$ (which has no physical effect,
since $|{\bf m}|=1$ at all points in the sample), and rotating the
coordinate system about the wire, $\sf K$~can be taken to equal ${\rm
diag}\,(-K_e,0,K_h)$ for some $K_e,K_h>0$.  The $z$-axis will be a hard
axis, and the $x$--$y$~plane an easy plane, within which the magnetization
vector at every point along the wire is preferentially confined.  A~hard
axis of this sort could arise from a shape anisotropy rather than a
crystalline anisotropy, e.g., from the cross section of the wire being
elliptical rather than circular.

For simplicity we consider here only the limiting case $K_h\to\infty$, in
which all fluctuational motion of the magnetization vector out of the
$x$--$y$~plane is suppressed.  (Taking out-of-easy-plane fluctuations into
account would alter the Kramers prefactor by multiplying it by an extra
determinant quotient, but would not affect the qualitative nature of its
$\epsilon\to0$ and $L\to\infty$ asymptotics.)  For this case the effective
Hamiltonian is
\begin{equation}
{\cal H}[{\bf m}]=
\int_0^L
\left\{
\left[
(\partial_x m_x)^2 + (\partial_x m_y)^2
\right]
-K_em_x^2 - h_{\rm ext}m_x
\right\}\,dx.
\end{equation}
As a functional of the angle~$\theta$ between~${\bf m}$ (in~the $x$--$y$
plane) and the $x$-axis, 
\begin{equation}
{\cal H}[\theta]=
\int_0^L
\left[
\left(\frac{d\theta}{dx}\right)^2 - K_e \cos^2\theta - h_{\rm ext}\cos\theta
\right]\,dx.
\end{equation}
In the absence of thermal noise, the stationary configurations of the
deterministic Landau--Lifshitz--Gilbert equation~(\ref{eq:llg2}) will be
those for which $\delta H[{\bf m}]/\delta{\bf m}=0$, i.e., $\delta
H[{\theta}]/\delta{\theta}=0$.  That~is, they will be those for which the
magnetization inclination angle~$\theta$ as a function of distance~$x$
along the wire satisfies
\begin{equation}
\label{eq:eulerlagrange}
2\frac{d^2\theta}{dx^2} + 2K_e\sin\theta\cos\theta + h_{\rm
ext}\sin\theta=0.
\end{equation}
For the moment, suppose that the wire can be viewed in some sense as
toroidal, i.e., suppose that periodic boundary conditions
$\theta(0)=\theta(L)$, $\theta'(0)=\theta'(L)$ can be imposed.  Then the
solutions of~(\ref{eq:eulerlagrange}) will include $\theta\equiv\pi$
and~$\theta\equiv0$, which are the familiar states of metastable and stable
magnetization, with the magnetization pointing against or along the
external field, which is itself applied parallel to the wire.  The solution
$\theta\equiv \cos^{-1}(-h_{\rm ext}/2K_e)$ is also stationary in~time,
though unstable.  This is the {\em uniform\/} (N\'eel--Brown) barrier to
magnetization reversal, which in this context is a saddle point in the
infinite-dimensional space of field configurations.

However, if $L$ is sufficiently large, there will also be a {\em
nonuniform\/} barrier to magnetization reversal.  By proceeding as in our
analysis~\cite{Maier02} of the Ginzburg--Landau equation~(\ref{eq:GL}), one
can show that
\begin{equation}
\label{eq:newelliptic}
\theta=\theta_u(x)\equiv
\cos^{-1}
\left(
\frac
{a_1\,{\rm sn}^2\left(\,(x-x_0)/c\,|\,m\,\right) + a_0}
{b_1\,{\rm sn}^2\left(\,(x-x_0)/c\,|\,m\,\right) + b_0}
\right),
\end{equation}
where $a_0,a_1,b_0,b_1,c,m$ depend on $L$ and $h_{\rm ext}/2K_e$, and
$x_0$~is arbitrary, is also an unstable stationary solution, and is at
lower energy than the N\'eel--Brown barrier.  Here ${\rm sn}(\cdot|m)$ is
the Jacobi elliptic function with modular parameter $m\in(0,1)$, the period
of which is~$4K(m)$, where $K(m)$~is the first elliptic
integral~\cite{Abramowitz65}.  The parameter~$m$ is of~course chosen so
that~$\theta_u$, as a function of~$x$, has period~$L$.  If~such a
lower-energy barrier to reversal exists, its `height' in~terms of energy,
which may be written as~$\Delta{\cal H}(L)$, will determine the exponential
falloff of the rate of escape from the metastable state $\theta\equiv\pi$,
as the noise strength tends to zero.

At this point, we have gone significantly beyond the analysis of Braun.
It~is not difficult to check that the solution $\theta=\theta_u(x)$ can be
viewed as a critical droplet: for many choices of parameter, it~resembles a
droplet of the $\theta=0$ phase inserted into a background consisting of
the $\theta=\pi$ phase, and delimited by a $\pi$~wall and a $-\pi$~wall.
However, its shape depends on the length~$L$ of the wire.  As~$L\to\infty$,
it tends to a limiting shape, which may be expressed in~terms of hyperbolic
trigonometric functions.  This is possible because as~$L\to\infty$, the
modular parameter~$m$ tends to unity, and ${\rm sn}(x|m)$ degenerates
to~$\tanh x$.  The formula~(\ref{eq:newelliptic}) is new in this context,
though the limiting shape as~$L\to\infty$ was obtained by Braun.  Similar
expressions for periodic micromagnetic structures, involving elliptic
functions, have been derived in other contexts~\cite{Shirobokov39}.

It follows immediately from the discussion in Section~\ref{sec:kramers}
that for this model of a thin cylindrical particle of length~$L$ with
periodic boundary conditions, the magnetization reversal rate~$\Gamma$
satisfies
\begin{equation}
\label{eq:novel2}
\begin{array}{ll}
\Gamma\,\sim\, C(L)\,\epsilon^{-1/2} \exp\bigl(-2(\Delta{\cal H})(L)/\epsilon\bigr),&\quad\epsilon\to0\,,\\[\jot]
C(L)\,\sim\,{\rm const}\times L,&\quad L\to\infty\,.
\end{array}
\end{equation}
The factor $\epsilon^{1/2}$ is due~to the fact that the separatrix between
the metastable and stable domains of attraction contains a limit cycle of
length~$L$, since the droplet may be formed anywhere along the wire, on
account of translation invariance.  ($x_0$~above is arbitrary.)  This is
also responsible for the asymptotic linear growth of the prefactor
with~$L$.  These two features were noted by Braun, but we now see they have
a simple physical origin.

The present approach, in which the shape of the critical droplet for
fixed~$L$ is precisely specified in~terms of elliptic functions, yields
precise values for the exponent $2\Delta{\cal H}(L)$ and the
prefactor~$C(L)$.  Details of the calculations, which are numerical rather
than analytic, may appear elsewhere.  What is of greater interest, however,
is the modification of this analysis to include more realistic boundary
conditions.  Braun has pointed~out that in a cylinder of finite length,
`open' boundary conditions with the magnetization perpendicular to the ends
may be appropriate~\cite{Braun99}.  In~the present scheme this corresponds
to taking $\theta'(0)=\theta'(L)=0$, i.e., to {\em Neumann\/} boundary
conditions.

Equipped with Neumann boundary conditions, the Euler--Lagrange
equation~(\ref{eq:eulerlagrange}) still has the metastable and stable
states $\theta\equiv\pi$, $\theta\equiv0$ as solutions.  A~detailed
analysis, which will be published elsewhere, reveals that by choosing
parameters appropriately, the unstable elliptic solution
$\theta=\theta_u(x)$ of~(\ref{eq:newelliptic}) can also be made to satisfy
Neumann boundary conditions.  However, when this is done, it becomes a
`half-droplet' delimited by a $\pi$~wall and either $x=0$ or~$x=L$.  Unlike
the case of periodic boundary conditions, nucleation at low temperatures is
therefore not likely to take place anywhere along the wire, but must, on
energetic grounds, preferentially begin at either end.  Once a critical
half-droplet has formed near either end, it may roll `downhill' in the
magnetization configuration space by spreading along the wire, replacing
the original value $\theta=\pi$ by $\theta=0$ as it goes.

The important point is that if Neumann rather than periodic boundary
conditions are imposed, there will be no infinite degeneracy of the
transition state, no~unstable limit cycle, no~non-Arrhenius
O($\epsilon^{-1/2}$) behavior of the Kramers prefactor, and also no
asymptotic linear dependence of the prefactor on~$L$ as~$L\to\infty$.  The
noise-activated magnetization reversal of a long thin ferromagnetic
cylinder actually fits much better into traditional Kramers theory if
Neumann boundary conditions are imposed.

\acknowledgments     
 
This research was supported in part by National Science Foundation Grant
No.~PHY-0099484.



\end{document}